\documentclass[12pt]{iopart}
\usepackage{graphicx}
\usepackage{iopams}
\usepackage{xcolor}

\begin{document}
%version 3 / \today

\title[FeGa$_{3-x}$Ge$_x$ studied by ESR]{Spin dynamics of FeGa$_{3-x}$Ge$_x$ studied by Electron Spin Resonance}

\author{Bonho Koo$^1$, Kristian Bader$^2$, Ulrich Burkhard$^1$, 
Michael Baenitz$^1$, Peter Gille$^2$ and J\"org Sichelschmidt$^1$}

\address{$^1$ Max Planck Institute for Chemical Physics of Solids, Dresden, Germany}
\address{$^2$ Ludwig-Maximilians-Universit\"at, Sektion Kristallographie, M\"unchen, Germany}
\ead{Sichelschmidt@cpfs.mpg.de}

%\author{Bonho Koo, Ulrich Burkhard, M. Baenitz, J\"org Sichelschmidt}
%
%\address{Max Planck Institute for Chemical Physics of Solids, 01187 Dresden, Germany}
%
%\author{Kristian Bader, Peter Gille}
%
%\address{Ludwig-Maximilians-Universit\"at, Sektion Kristallographie, M\"unchen, Germany}

\begin{abstract}
The intermetallic semiconductor FeGa$_{3}$ acquires itinerant ferromagnetism upon electron doping by a partial replacement of Ga with Ge. We studied the electron spin resonance (ESR) of high-quality single crystals of FeGa$_{3-x}$Ge$_x$ for $x$ from 0 up to 0.162 where ferromagnetic order is observed. For $x = 0$ we observed a well-defined ESR signal, indicating the presence of pre-formed magnetic moments in the semiconducting phase. Upon Ge doping the occurrence of  itinerant magnetism clearly affects the ESR properties below $\approx 40$~K whereas at higher temperatures an ESR signal as seen in FeGa$_{3}$ prevails independent on the Ge-content. The present results show that the ESR of FeGa$_{3-x}$Ge$_x$ is an appropriate and direct tool to investigate the evolution of 3d-based itinerant magnetism.

%Fe-based intermetallic semiconductors such as FeSi, FeSb$_{2}$, and FeGa$_{3}$ are known to be diamagnetic due to the strong hybridization between the Fe d levels and s,p levels. From this diamagnetic behavior, naturally, no electron spin resonance (ESR) can be expected. However, the particular single crystal of FeGa$_{3}$ we have investigated shows a clear and well-defined ESR signal. A partial replacement of Ga-sites with Ge leads to metallic behavior, the formation of local magnetic moments, and ferromagnetic order. This transition by Ge-doping is reflected in strong changes of the low-temperature properties of the ESR signal. For FeGa$_{3-x}$Ge$_x$, $x=0.127$ (high-quality single crystal), the linewidth and the electrical resistivity follow the same temperature dependence. This indicates that the observed ESR signal is due to the resonance of conduction electron spins. The present results show that the ESR of FeGa$_{3-x}$Ge$_x$ is an appropriate and direct tool to investigate the evolution of ferromagnetic correlations in the vicinity of a ferromagnetic quantum critical point.
%
\end{abstract}

\section{Introduction}

% questions to be discussed with ESR results:
% - in the framework of investigating FM QC: evolution of magnetic order with Ge doping - the role of electronic phase separation - or magnetic clusters?
% - ESR related: which relaxation mechanism? Examples of other FM metals. 
% - FeGaGe related: is magnetism comparable to materials without magnetic element - such as ZrZn2, Sc3In,...? Here, the detailed structure of the DOS at Ef is important.

FeGa$_{3}$ is expected to be a nonmagnetic, narrow-gap semiconductor with a gap of $\approx0.4$~eV that originates from a strong hybridization of the Fe d and Ga p atomic orbitals \cite{yin10a,arita11a}. Doping with Ge leads to an evolution of a metallic magnetic state which for low Ge substitutions shows properties of an antiferromagnetic $3d$-electron heavy fermion system. At a critical Ge-concentration of $x=0.15$ ferromagnetic (FM) order sets in and three-dimensional quantum critical FM fluctuations are indicated by $^{71}$Ga nuclear quadrupole resonance (NQR) relaxation data \cite{majumder16a}. FeGa$_{3-x}$Ge$_x$ serves as a rare platform in which $3d$ heavy fermions occur in the vicinity of a FM quantum critical point with pronounced three-dimensional FM fluctuations.\\
%
%As illustrated in Fig. \ref{FeGa3GePD} ferromagnetic (FM ) order is created for a critical Ge-concentration of $x=0.15$. For this concentration three-dimensional (3D) quantum critical FM fluctuations are indicated by $^{71}$Ga NQR relaxation data \cite{majumder16a}. These NQR results also indicate the absence of strong intrinsic disorder due to Ge doping. Hence, FeGa$_{3-x}$Ge$_x$ serves as a rare model system in which $3d$ heavy fermions occur in the vicinity of a quantum critical point with pronounced 3D FM fluctuations.\\
%%%%%%%%%%%%%%%%%%%%%%%%%%%%%%%%%%%%%%%%%%%%%%%%
%\begin{figure}[h]
%\begin{center}
%\includegraphics[width=0.6\columnwidth]{FeGa3GePD}
%\end{center}
%\caption{
%Phase diagram of FeGa$_{3-x}$Ge$_x$ illustrating heavy fermion (HF) behavior below the critical concentration $x=0.15$ above which short-range and long-range ferromagnetic order (FM SRO and FM LRO) is found \cite{majumder16a,PD}. Ferromagnetic quantum critical (FMQC) fluctuations are indicated by the $^{71}$Ga NQR spin-lattice relaxation $1/TT_{1}$ which contains transferred magnetic spin fluctuations of Fe-$3d$ spins, $R_{3d}$, and conduction electrons, $R_{CE}$. $n$ denotes the exponent in a power law $AT^{-n}$ which describes $R_{3d}$.\cite{majumder16a}
%}
%\label{FeGa3GePD}
%\end{figure}
%%%%%%%%%%%%%%%%%%%%%%%%%%%%%%%%%%%%%%%%%%%%%%%%
%%
Yet, the nature and evolution of the magnetic order is not clear. Models with itinerant magnetism as well as the local moment scenario are discussed. Density functional theory (DFT) calculations show that itinerant  ferromagnetism (possibility a Stoner mechanism) can be produced by electron- (or hole-) doping of FeGa$_{3}$ (independent on the presence of pre-formed moments in FeGa$_{3}$) \cite{singh13b}.
However, the magnetic moments are not evenly distributed throughout all Fe atoms as expected in an itinerant picture. The magnetic moments introduced by Ge are found to be distributed on the Fe atoms in a complex, strongly Ge-concentration dependent way and it is under strong discussion if all Fe moments are itinerant or if some stay localized \cite{alvarez-quiceno16a}. This may also explain that the mainly ferromagnetic magnetic ordering display a minor AFM component which leads to clear differences in field cooled / zero field cooled magnetization.
% electronic phase separation: ``self-trapping'' of doped electrons in a lattice distortion
Therefore, experimental methods sensitive to the local Fe environment are desirable. Muon spin rotation ($\mu$SR) and M\"ossbauer spectrocopies provided evidence for a continuous evolution of magnetic order from short-range below and near the critical concentration to long-range for higher Ge-concentrations \cite{munevar17a}. Moreover, the $\mu$SR data show that the distribution of the Ge on the two Ga sites in this host strongly determines the character of the magnetic order, confirming the results of DFT calculations \cite{alvarez-quiceno16a}.

The technique of electron Spin Resonance (ESR) is like NQR another microscopic probe for magnetism. It has proven to provide direct access to the spin dynamics of many ferromagnetic metals, i.e. itinerant magnets where the magnetism arises from conduction electrons.
In general, a conduction electron spin resonance (CESR) in itinerant magnets shows a narrow and observable linewidth in the presence of ferromagnetic correlations. This holds, in particular, for Kondo lattice systems like YbRh$_{2}$Si$_{2}$ and CeRuPO \cite{krellner08a} but also for 3d itinerant magnets such as ZrZn$_{2}$ \cite{walsh70a,forster10a}, Sc$_{3.12}$In \cite{shaltiel88a}, TiBe$_{2}$ \cite{shaltiel80a}, or NbFe$_{2}$ \cite{forster10a,rauch15a}. Recently, the interplay between a CESR and ferromagnetism was investigated for Cr$_{2}$B where a paramagnetic to ferromagnetic transition is induced by Fe-doping \cite{arcon16a}.
Without ferromagnetic correlations the CESR linewidth is mainly determined by spin-flip scattering which also determines the electrical resistivity. 

In this paper we could interpret the spin resonance in FeGa$_{3-x}$Ge$_x$ in a dual way: at low temperatures and $x>0.06$ in terms of a CESR with strong ferromagnetic correlations and at high temperatures in terms of a resonance of local spins.

%Examples of ESR in Fe-containing correlated materials: FeSe \cite{zhang13a}, Fe-Pnictides \cite{wu09a}, Fe3O4 (Verwey charge order transition near 120~K)
%
%A joint ESR and NMR investigation FeSe$_{0.42}$Te$_{0.58}$ suggested the coexistence of localized and itinerant electronic states \cite{arcon10a}. These results are discussed to be consistent with intraband electronic correlations that lead to a localization of one of the Fe-derived bands.
%
%magnetically ``disorder'' should be characterized by INS
%
%Neutron powder diffraction experiments suggested that FeGa$_{3}$ is an antiferromagnet with the ordering temperature $T_{\rm N}$ above 300~K and with ordered moments less than $1.5\mu_{\rm B}$ per Fe \cite{gamza14a}.
%
%Small-angle neutron scattering (D. Sokolov, Feb. 2017) on x=0.16 for q=0: almost no scattering, i.e. very small effective magnetic moment.\\

%
  %FeGa$_{3}$ is a local antiferromagnet
% wherein FeGa3:Ge is considered to be  How FM is created? $T_{\rm K}~1$~K, no local moment signatures in FeGa3 \cite{majumder16a}. few ppm Fe impurities in $\chi$\\
%
%M\"o\ss bauer (K. Bader): typical quadrupole splitting observed; fit the spectra by multiple Fe Species (Fe2+,Fe3+); 
%Can high-/ low-spin Fe be identified by  M\"o\ss bauer?? 
%
%homogeneity issue: WDX investigations (U. Burkhard)

%
\section{Experimental}
%\subsection{ESR}
Electron Spin Resonance (ESR) probes the absorbed power $P$ of a transversal magnetic microwave field as a function of an external magnetic field $\mu_0H=B$. To improve the signal-to-noise ratio, a lock-in technique is used by modulating the static field, which yields the derivative of the resonance signal d$P$/d$B$. The ESR experiments were performed at X- and Q-band frequencies ($\nu $=9.4 GHz and 34~GHz) using a continuous-wave ESR spectrometer. 
The sample temperature was set with a helium-flow cryostat allowing for temperatures between 2.7 and 300~K and a nitrogen-gas flow cryostat for temperatures up to 500~K.
%\emph{[L-band (1 GHz) and Q-band (34 GHz) frequencies are also available and will be used to provide a substantial data basis for the ESR in FeGa$_{3-x}$Ge$_x$]}

The obtained spectra were fitted by a Lorentzian line shape yielding the parameters linewidth $\Delta B$ which is a measure of the spin-probe relaxation rate, and resonance field $B_{\rm res}$ which is determined by the effective $g$-factor $(g=h\nu/\mu_{\rm B}B_{\rm res}$) and internal fields. 
The fitting included also a dispersion-to-absorption ratio $D/A$ from metallic contributions, as was done for the ESR spectra in the ferromagnetic metal NbFe$_{2}$ \cite{rauch15a}. $D/A$ parametrizes an asymmetry in the lineshape. $D/A=1$ refers to ESR spins within the microwave skin depth which is smaller than the sample thickness due to the electrical conductivity.  
The ESR intensity $I_{ESR}$ corresponds to the integrated ESR absorption and was calculated using the line amplitude, line width and D/A as reported in reference \cite{wykhoff07b}. $I_{ESR}$ is determined by the static spin-probe susceptibility, i.e. it provides a microscopic probe of the sample magnetization.

%\subsection{Sample preparation and characterization}

%\emph{As the first investigations of on FeGa$_{3-x}$Ge$_x$ were carried out exclusively on powder there was a strong interest in highly pure single crystals to extend the applied methods (resistivity, Hall, thermopower). Furthermore, there was need to improve the quality of the material (Fe-containing impurities) for a perfect crystal growth. Prof. Peter Gille (LMU Munich) succeeded to synthesize ultrapure and large single crystals. Large crystals of FeGa$_{3}$ are in particular suitable for planned neutron scattering studies (O. Stockert, D. Sokolov) to clarify the magnetic ground state.}

%concerning ``disorder'' in Ge-doped FeGa3: NQR does not necessarily provide the information on ``disorder'' which is relevant for ESR. Electronic phase separation needs to be considered!

% chemically ``disorder'' needs to be checked -- U. Burkhard (EDX) (see Baenitz talk: WDX by Peter Gille RetreatHarnackhaus2016-FeGa3 for Uli-Raul-Baenitz.pdf)

%magnetically ``disorder'' should be characterized by INS

%High-quality, ultrapure single crystals of FeGa$_{3}$ and FeGa$_{3-x}$Ge$_x$  were grown by the Czochralski method \cite{wagner-reetz14c}. Wavelength dispersive X-ray spectroscopy on single crystalline FeGa$_{3}$ demonstrated a homogenous sample constitution.

%The ESR experiments on FeGa$_{3-x}$Ge$_x$ were performed on pieces cut from one large Czochralski-grown crystal which shows a Ge-concentration gradient. This gradient was determined by ... We investigated $x=0, 0.06, 0.127, 0.133, 0.162$.

% P.Gille:
High-quality, ultrapure single crystals of FeGa$_{3}$ and FeGa$_{3-x}$Ge$_x$ were grown by the Czochralski method from Ga-rich solutions at temperatures below the peritectic transformation. Whereas for FeGa$_{3}$ the crystal growth has already been described elsewhere \cite{wagner-reetz14c} the problem with FeGa$_{3-x}$Ge$_x$ was the missing ternary Fe-Ga-Ge phase diagram. We assumed an only weak change of the liquidus by adding small amounts of Ge (bulk, 99.999\%, ChemPur) and did a couple of normal freezing experiments in order to determine the solid-phase compositions being in equilibrium with various ternary Ga-rich solutions. From this, we found a pseudobinary segregation coefficient $<1$ resulting in a slight axial increase of the Ge content along the growth direction. Starting in three Czochralski growth runs with ternary solutions in which 2.35, 5.0, and 5.75 \% of the Ga was substituted with Ge, we got FeGa$_{3-x}$Ge$_x$ single crystals in the range $0.054 \le x \le 0.165$. All Czochralski-grown crystals were pulled along the [001] direction using undoped FeGa$_{3}$ seeds and very low pulling rates of 0.05 to 0.1 mm/h.
The ESR experiments on FeGa$_{3-x}$Ge$_x$ were performed on pieces cut from these crystals showing an axial Ge-concentration gradient. This gradient was determined by electron probe microanalysis (WDX). We studied samples with x = 0, 0.06, 0.127, 0.133, and 0.162.

\section{Results}
The observed ESR in FeGa$_{3-x}$Ge$_x$ shows two different and independent types of spectra, one with and one without dispersive (i.e. metallic) contribution. The metallic type appears only for higher Ge concentrations ($x>0.06$) and only for temperatures below $\approx 40$~K. The other type of signal is very weak (compared to the first type) and is seen above $\approx 40$~K (for $x=0$ above 5~K) up to the highest investigated temperature of 500~K. The presence of these two different spectra types indicates the complex way magnetism is created in FeGa$_{3-x}$Ge$_x$ which is a topic that recently came into focus \cite{alvarez-quiceno16a}. In the following we present and discuss the ESR spectra for undoped FeGa$_{3}$ and the effect of Ge doping on the spectra.  

%%%%%%%%%%%%%%%%%%%%%%%%%%%%%%%%%%%
\subsection{FeGa$_{3}$}

\begin{figure}[hbt]
\begin{center}
\includegraphics[width=0.8\columnwidth]{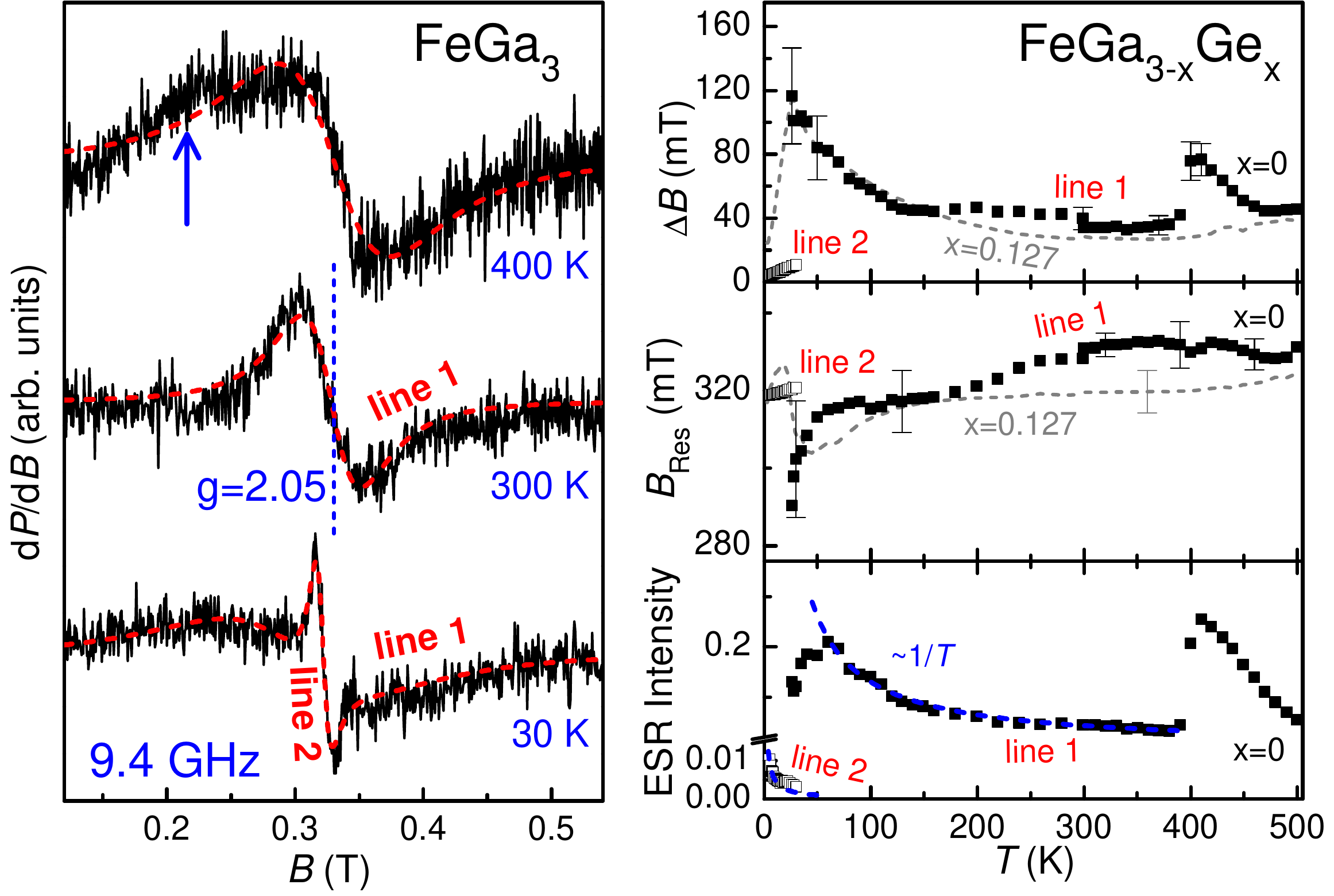}
\end{center}
\caption{
Left: X-band (9.4~GHz) ESR spectra (scaled to common amplitudes) of FeGa$_{3}$ at temperatures $T$ as indicated. Dashed lines denote a Lorentzian line fit. Arrow indicates an additional structure at $T\gtrsim400$~K which causes an effectively larger linewidth of the fitted line.  
Right: Temperature dependence of the ESR line parameters linewidth $\Delta B$, resonance field $B_{\rm res}$ and ESR intensity (being largely consistent with a Curie law, blue dashed line). Dashed grey lines indicate for comparison the parameters for x=0.127. Labels ``line1'' and ``line 2'' refer to the parameters of the corresponding lines in the left frame.
}
\label{SpecsPara2b}
\end{figure}

Typical ESR spectra in the high and low temperature regime are shown in the left frame of figure \ref{SpecsPara2b} together with fits by Lorentzian lines which result in temperature dependent parameters as shown in the right frame of figure \ref{SpecsPara2b}. The spectra consist of two well-defined resonance lines, a broad and a narrow one, denoted as line 1 and line 2. The narrow line appears at temperatures below $\approx 40$K whereas the broad one is observable in the whole temperature range 5-300~K. Both lines appear at resonance fields corresponding to a $g$-value of 2.05 This indicates a configuration 3$d^{5}$ (Fe$^{3+}$)  rather than 3$d^{6}$ (Fe$^{2+}$) for which, due to contributions from the orbital momentum, a much larger g-value is expected \cite{abragam70a}. 

We interpret the narrow line 2 as a resonance of Fe-impurity centers because its intensity is very small ($\approx 1000$ times weaker than line 1, note the intensity scale in figure \ref{SpecsPara2b}) and line 2 is not reproducible in samples from different batches.

Regarding line 1 it is important to note that very similar lines are also observed in the Ge-containing FeGa$_{3}$ (see the following section). This particularly holds for the temperature dependencies of all ESR parameters as shown in the right frame of figure  \ref{SpecsPara2b} where the dashed lines refer to the Ge-concentration x=0.127. For both x=0 and x=0.127 (and all the other Ge-concentrations, see figure \ref{FeGa3GeSpecsPara}) a continuous increase of $\Delta B$ and decrease of $B_{\rm res}$ is observed towards low temperatures (discussion in the following paragraph). The ESR intensity shows a clear Curie law $I^{ESR}_{1}\propto 1/T$ which indicates the local character of the ESR-active probes.

At high temperatures a remarkable kink in linewidth and intensity occurs. It originates from an additional line superimposed on line 1 as indicated by the blue arrow in figure \ref{SpecsPara2b}. Near above $\approx 400$~K this line dominates the single-Lorentzian line fitting. With further heating this additional line gets narrower and less intense which both is characteristic for paramagnetic centers near magnetic order. Thus the appearance of this spectral feature suggests the onset of magnetic order near below 400~K of some of the paramagnetic centers. This conclusion is consistent with neutron diffraction measurements which indicate antiferromagnetic order in FeGa$_{3}$ above 300~K \cite{gamza14a}. The presence of Fe-rich foreign phases like Fe$_{3}$Ga$_{4}$ with its magnetic transition at 360~K \cite{mendez15a} should also be taken into account to understand features at high temperatures.

%/ info from Hiroshi Yasuoka and Yasuki Kishimoto:
%Nuclear quadrupole resonance (NQR) spectra of $^{69}$Ga are characterized by two relaxation times $T_{1}$: short (~0.4s) and long (~36s). Both of these components may be related to the two ESR lines. Assuming that the short $T_{1}$ corresponds to the narrow ESR line then exchange narrowing would dominate the linewidth of line 2 - i.e. the narrowing is created by fast fluctuations which determine the short $T_{1}$. However, the NQR relaxation rates show no clear temperature dependences (measurements done only at 4.2, 20, 40, and 70~K), and, hence, the way how the ESR and NQR relaxation mechanism are related remains unclear. 
%The main broad line (``line 1'') remains symmetric without any dispersion contribution. This means the low conductivity of the sample allows the microwave to penetrate the whole sample volume (about 0.5x1x2mm$^{3}$ - needs to be checked). 

%%%%%%%%%%%%%%%%%%%%%%%%%%%%%%%%%%%%%%
\subsection{FeGa$_{3-x}$Ge$_x$}

\begin{figure}[h]
\begin{center}
\includegraphics[width=0.8\columnwidth]{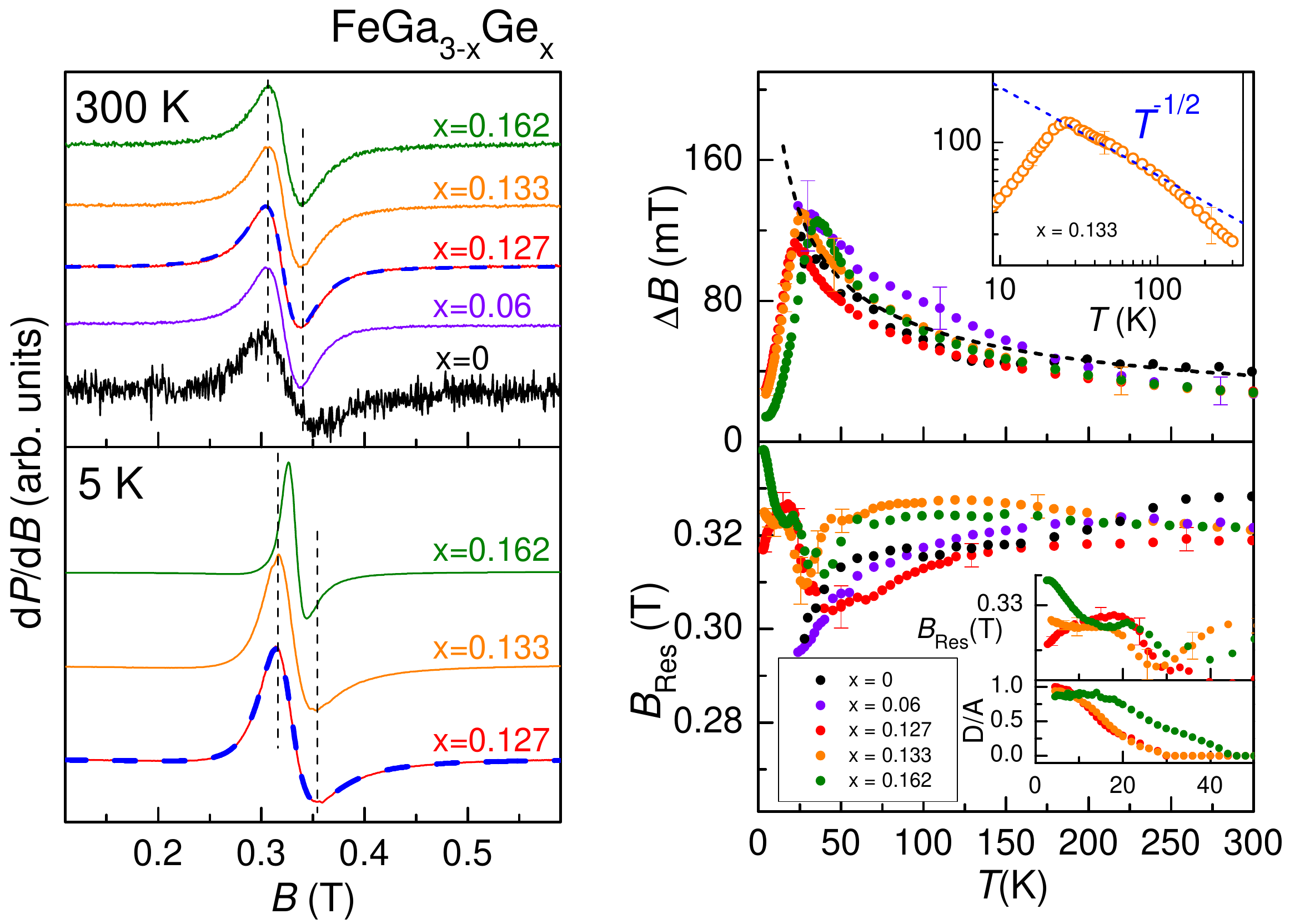}
\caption{
Left: Rescaled ESR spectra of FeGa$_{3-x}$Ge$_{x}$ with various x as indicated at 5~K and 300~K. For x = 0 and 0.06 no resonance was observable at 5 K. Dashed lines refer to Lorentzian line fittings. Note the asymmetric shape at 5~K indicative for a conductive environment. Right: temperature dependence of linewidth $\Delta B$ and resonance field $B_{\rm res}$. Dashed lines in upper frames indicates $T^{-0.5}$ behavior. Inset in lower frame shows $B_{\rm res}$ together with the line asymmetry parameter D/A. 
}
\label{FeGa3GeSpecsPara}
\end{center}
\end{figure}

A partial substitution of Ga by Ge leads to a strong change in the magnetic properties of FeGa$_{3}$ (see the phase diagrams in Refs. \cite{majumder16a,munevar17a}). 
Also, as illustrated in figure  \ref{FeGa3GeSpecsPara}, the ESR properties change dramatically, namely
for the higher concentrations $x\ge0.127$ a striking change appears in linewidth and resonance field at temperatures of $\approx 40$K.  
When cooling below this temperature we observed a remarkable crossover to a strong and narrow line developing out of the noisy and relatively weak high-temperature line. This crossover is most obviously seen in the linewidth but also, as shown in the lower insets of figure  \ref{FeGa3GeSpecsPara}, in the resonance field and the dispersion-to-absorption ratio (D/A). The change in resonance field is strongest for x=0.162 which is the composition with the strongest ferromagnetic correlations \cite{majumder16a,munevar17a}. The change from a symmetric signal (D/A=0) to an asymmetric signal (D/A=1) demonstrates that the influence of the electrical conduction in the environment of the spin probe becomes important. Again it is the sample with the strongest ferromagnetic correlations, x=0.162, where a pronounced line asymmetry reaches up to the highest temperatures.  

Above $\approx 40$K the ESR lines and their properties seem to remain unchanged by Ge doping (compare Figs. \ref{SpecsPara2b} and \ref{FeGa3GeSpecsPara} and note the common plot of parameters for x=0 (solid square) and x=0.127 (dashed line) in figure  \ref{SpecsPara2b}). With increasing temperature the linewidth continuously decreases in way agreeing well with a $T^{-1/2}$ law (dashed line in figure  \ref{FeGa3GeSpecsPara}). This behavior points towards a relaxation by spin fluctuations if they have a spin correlation rate $\propto T^{1/2}$ \cite{cox87a} and the static susceptibility showing a local Curie law. A variation of this powerlaw behavior among the various Ge contents should be expected on the basis on NQR results on the spin correlation rate \cite{majumder16a}. However, such a variation could not be resolved by our ESR linewidth data.

%The origin of the increase of $\Delta B$ with decreasing temperature may also be due to ferromagnetic correlations. Usually, in systems with magnetic exchange coupling approaching magnetic order leads to a reduction of the linewidth exchange narrowing. \\
Regarding the weak temperature dependence of the resonance field $B_{\rm res}$ above $\approx 40~K$ one might relate its origin to the temperature evolution of effective internal magnetic fields. The presence of such internal fields sounds most plausible for FeGa$_{3}$ $(x=0)$ for which AFM order is suggested by our high-temperature ESR results (see section below).

%The continuous decrease of $B_{\rm res}$ with decreasing temperature indicates the growth of internal magnetic fields in the sample, similar to the ESR of Fe-doped Cr$_{2}$B which shows ferromagnetic correlations \cite{arcon16a}.

\begin{figure}[h]
\begin{center}
\includegraphics[width=0.6\columnwidth]{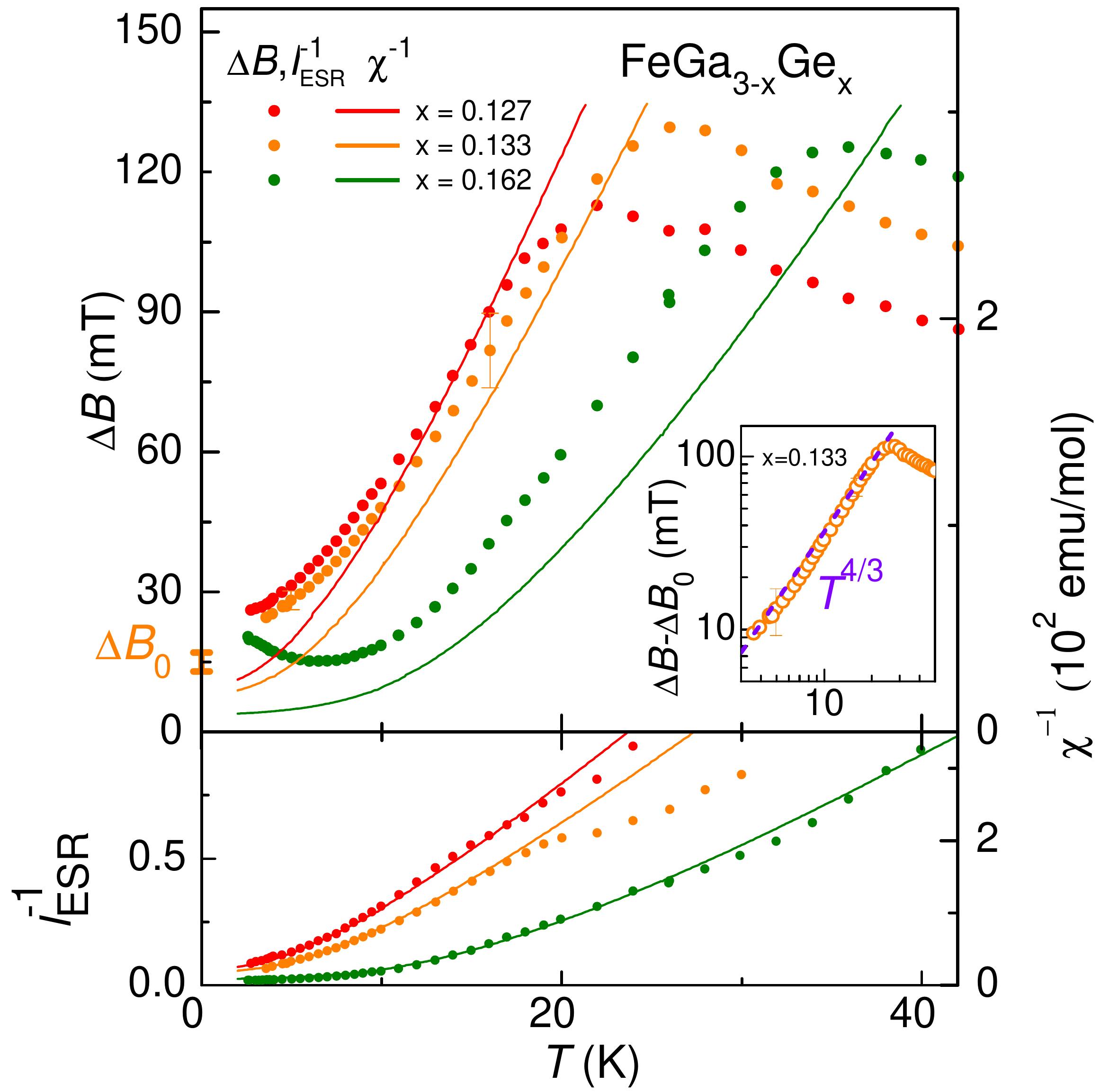}
\caption{
Low-temperature properties of the X-band ESR linewidth $\Delta B$, inverse intensity $I_{ESR}^{-1}$ and magnetic susceptibility $\chi^{-1}$ (solid lines, for 0.33~T at same samples). Inset highlights a power law in the temperature dependence of the reduced linewidth for x=0.133 where a residual $\Delta B_{0}=15$~mT as indicated in the main frame was used.
}
\label{FeGa3GedHChiInt}
\end{center}
\end{figure}

For the compositions with $x\ge0.127$ the low-temperature ESR results are influenced by strong FM correlations and FM ordering. This is illustrated in greater detail in figure  \ref{FeGa3GedHChiInt} where the linewidth and ESR intensity is compared with the DC magnetic susceptibility $\chi$ measured at the same samples. Broad linewidth maxima indicate a crossover to a low-temperature region with an intense, narrow and  asymmetric ESR.  As shown in the lower frame the temperature dependencies of both ESR intensity and susceptibility coincide roughly below temperatures where the linewidth displays a broad maximum. This demonstrates that in the low-temperature region the ESR probes the same bulk magnetic properties that determine the DC magnetic susceptibility.

\begin{figure}[h]
\begin{center}
\includegraphics[width=0.6\columnwidth]{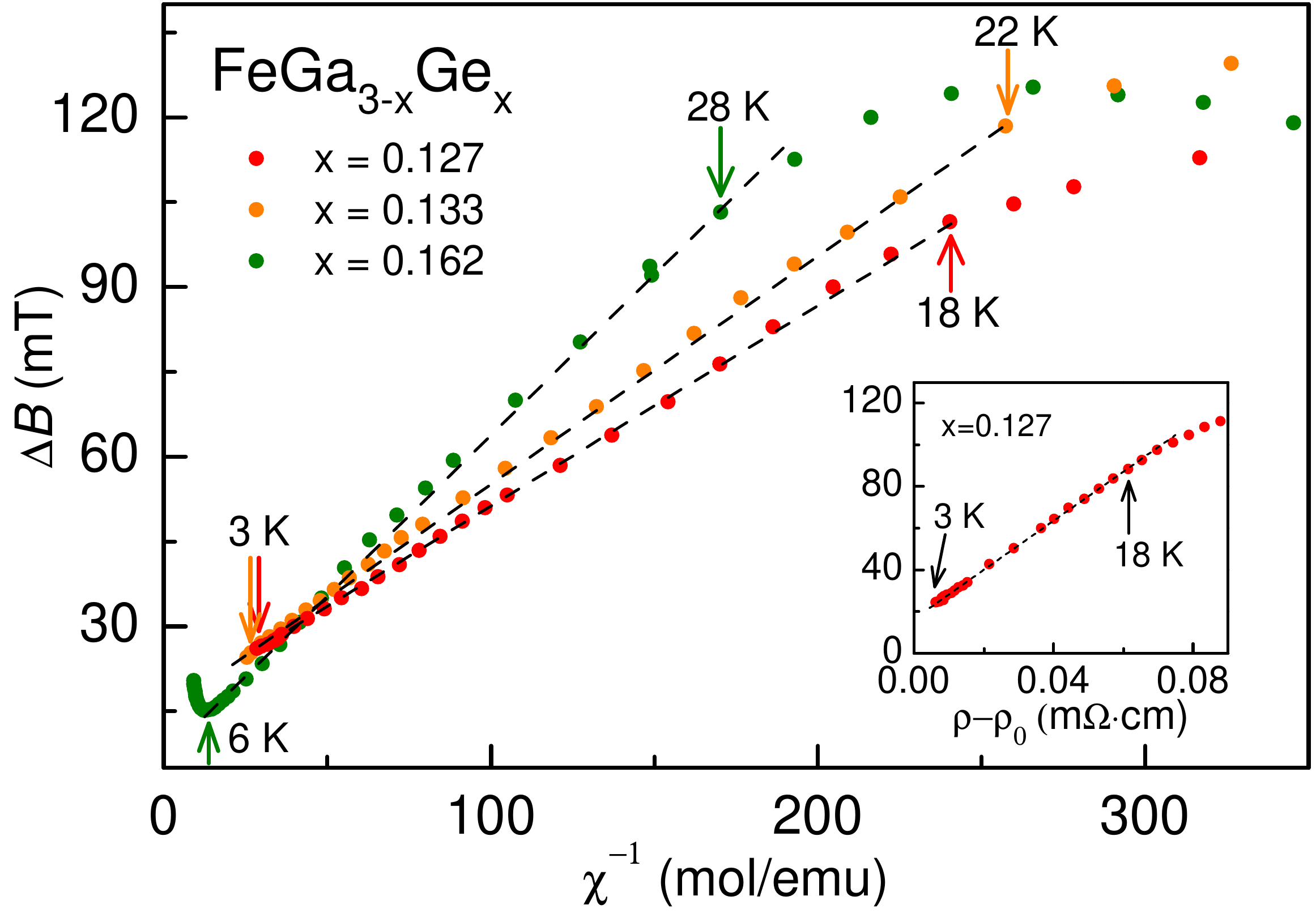}
\end{center}
\caption{
Relation of the X-band linewidth $\Delta B$ (at $T<40$~K) to the inverse magnetic susceptibility $\chi^{-1}$ and the electrical resistivity $\rho-\rho_{0}$ with temperature as implicit parameter ($\rho_{0}=0.7159\;\mathrm{m\Omega cm}$ denotes the temperature independent part of $\rho(T)$). Dashed lines indicate linear behavior.
}
\label{FeGa3GedHRhoChi}
\end{figure}
Comparing the linewidth with the reciprocal susceptibility as shown in the upper frame of figure  \ref{FeGa3GedHChiInt} suggests a direct relation between these quantities. Plotting $\Delta B$ vs. $\chi^{-1}$ as shown in figure  \ref{FeGa3GedHRhoChi} further establishes such a relation which in turn is consistent with the properties of a CESR in spin systems with magnetic correlations. As was observed, for instance, in the itinerant ferromagnet ZrZn$_{2}$ \cite{walsh70a}, a 'ferromagnetic slowing down' of the relaxation rate (or  'exchange enhancement' of the spin lifetime) \cite{walsh70a,barnes81a} strongly reduces the spin-orbit relaxation und thus leads to a narrow CESR linewidth with $\Delta B\propto \chi^{-1}$.
%the internal or exchange field impedes the spin relaxation resulting in a narrowing of the CESR linewidth, also called exchange enhancement of the spin lifetime \cite{walsh70a}.
This relation allows to check whether a power law behavior in $\chi\propto T^{-4/3}$ for three-dimensional ferromagnetic quantum critical fluctuations \cite{mishra98a} can be found in the ESR linewidth. Indeed, as shown in the inset of figure  \ref{FeGa3GedHChiInt}, taking into account a reasonable residual, temperature independent contribution $\Delta B_{0}$, the linewidth temperature dependence for $x=0.133$ is consistent with $T^{4/3}$. This power law is also reported for the NQR-probed dynamical susceptibility of the critical composition $x=0.15$ \cite{majumder16a}.

%At the same time, the lineshape acquires a pronounced dispersion contribution, the D/A ratio approaches one, and a clear asymmetry appears in the lineshape (left frame and middle of the right frame of Fig. \ref{FeGa3GedHChiInt}). Such a behavior points towards a decrease of microwave penetration depth and a concomitant increase of conductivity towards low temperatures.\\

The inset of figure  \ref{FeGa3GedHRhoChi} illustrates for x=0.127 a characteristic hallmark of a CESR, namely a linear relation between the temperature dependencies of linewidth and electrical resistivity ($\rho-\rho_{0} \propto T^{5/3}$). This behavior is similarly observed for the itinerant ferromagnets ZrZn$_{2}$ and NbFe$_{2}$ above magnetic ordering \cite{forster10a,rauch15a}.
%Note that the residual resistivity $\rho_{0}$ has a relatively large value when compared with $\rho-\rho_{0}$ \emph{(how conclusive is the $T^{5/3}$-behavior then?)}, whereas the temperature dependent effects in the linewidth are more pronounced. 
%
%

\begin{figure}[h]
\begin{center}
\includegraphics[width=0.5\columnwidth]{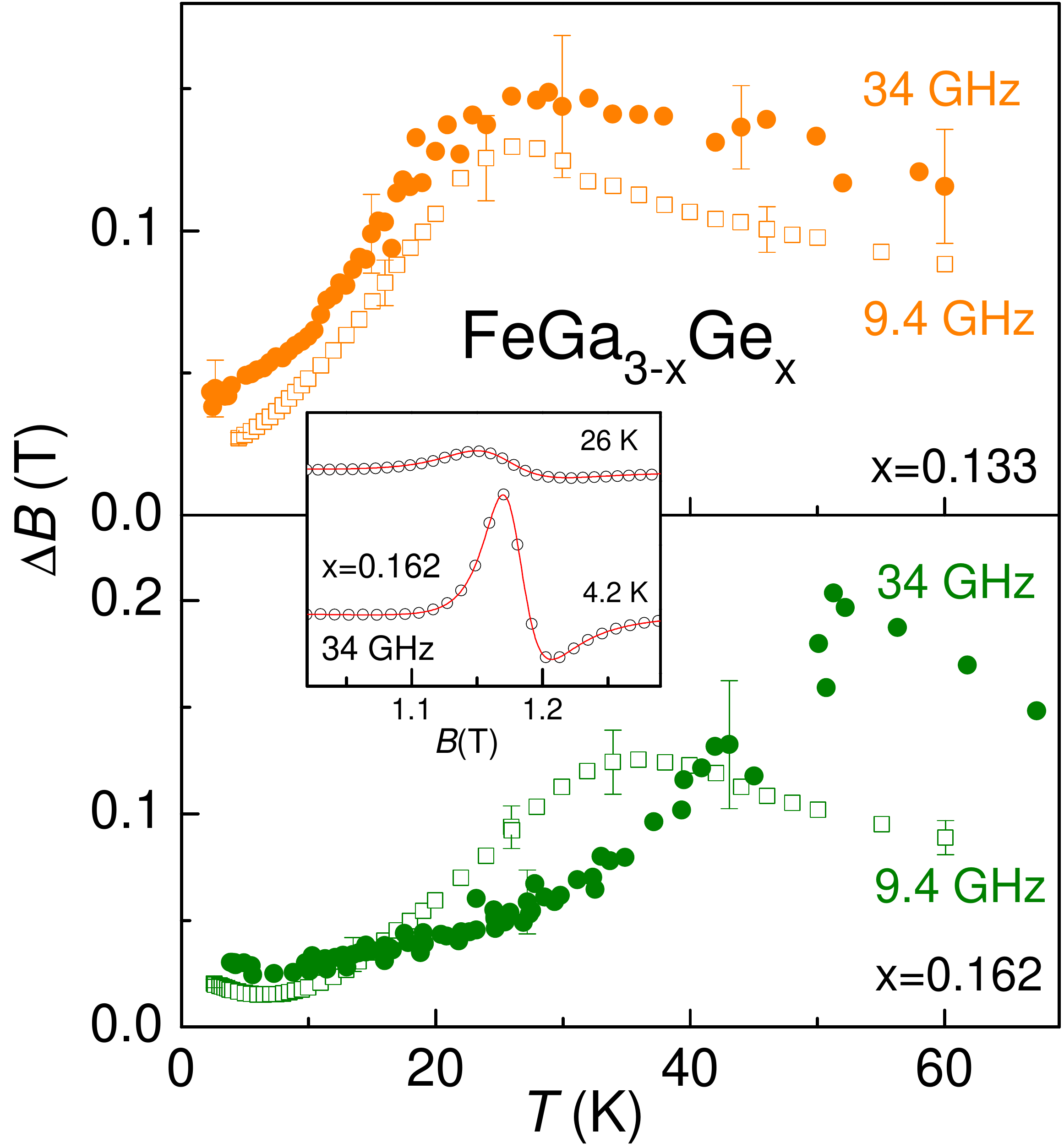}
\end{center}
\caption{
Comparison of the temperature dependence of the ESR linewidth $\Delta B$ measured for 9.4~GHz (X-band) and 34~GHz (Q-band). Inset shows selected Q-band spectra (open circles) together with a Lorentzian fit (solid line) of their lineshapes. 
}
\label{FeGa3GeQband}
\end{figure}

The above discussed presence of an 'exchange enhancement' implies that the influence of ferromagnetic correlations on the ESR linewidth should depend on the applied magnetic field. The so far presented ESR data were taken at $\nu=9.4$~GHz (X-band) that corresponds to a resonance at a field of $\approx0.31$~T (figure  \ref{FeGa3GeSpecsPara}). Using $\nu=34$~GHz (Q-band) with $B_{\rm res}\approx 1.17$~T one can expect a strengthened effect of the ferromagnetic correlations to the ESR parameters. This, indeed, could be observed as shown in figure  \ref{FeGa3GeQband}. The low-temperature linewidth depends on both the Ge content and the resonance field (at X-, or Q-band). The stronger the ferromagnetic correlations, i.e. the larger the Ge content, the more clear are the effects on the linewidth.

\begin{figure}[h]
\begin{center}
\includegraphics[width=0.4\columnwidth]{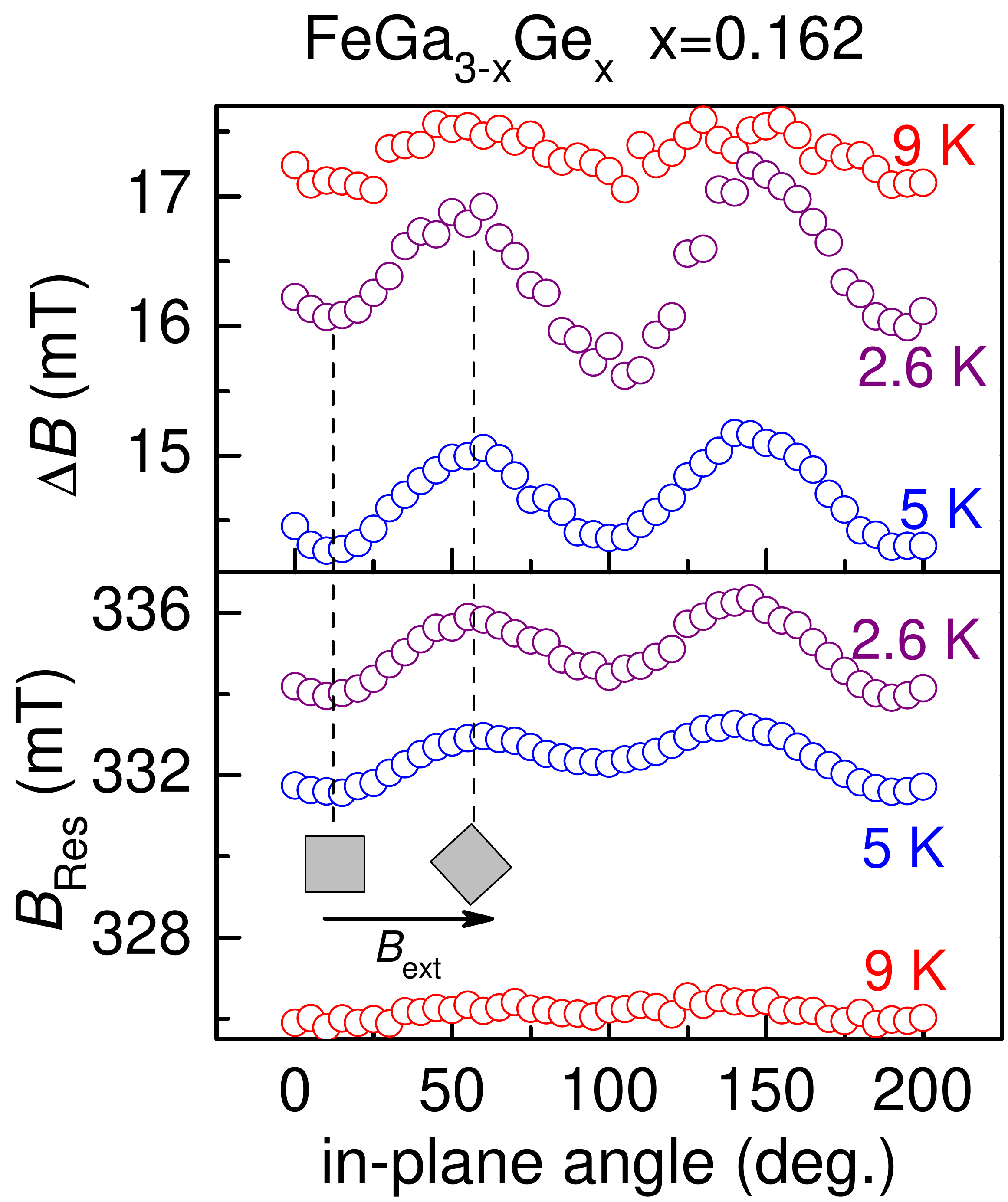}
\end{center}
\caption{
In-plane angle dependence of X-band linewidth $\Delta B$ and resonance field $B_{\mathrm Res}$ on the sample rotation around the crystallographic c-axis with the applied field $B\perp c$. At low temperatures a weak anisotropy indicates magnetic order. The field-orientation of the base of the cuboid shaped sample is shown for two angles by the grey squares.
}
\label{FeGa3GeAngle}
\end{figure}

The linewidth is quite obviously related to the Ge content (see figure  \ref{FeGa3GedHChiInt}) which determines the formation of ferromagnetism in a delicate way. Muon spin rotation results indicate short range ferromagnetic order for a Ge content larger than $x=0.11$ \cite{munevar17a}. The weak  increase of the linewidth for $x=0.162$ at low temperatures evidences long-range ferromagnetic order which leads to pronounced anisotropies in the ESR parameters as shown in figure  \ref{FeGa3GeAngle}. The crystal was rotated around the crystallographic $c$-axis and a clear anisotropy with a 90 degree periodicity was found only for temperatures below the ferromagnetic ordering temperature $T_{\rm C}=6$~K. M\"ossbauer spectroscopy results \cite{munevar17a} have demonstrated that the ordered moments are aligned in the plane perpendicular to the crystallographic $c$-axis. Thus, the observed anisotropies shown in figure  \ref{FeGa3GeAngle} are indeed a consequence of FM magnetic ordering and point out that the observed line actually is a ferromagnetic resonance mode.  
\subsection{High temperature properties.}
\begin{figure}[h]
\begin{center}
\includegraphics[width=1.0\columnwidth]{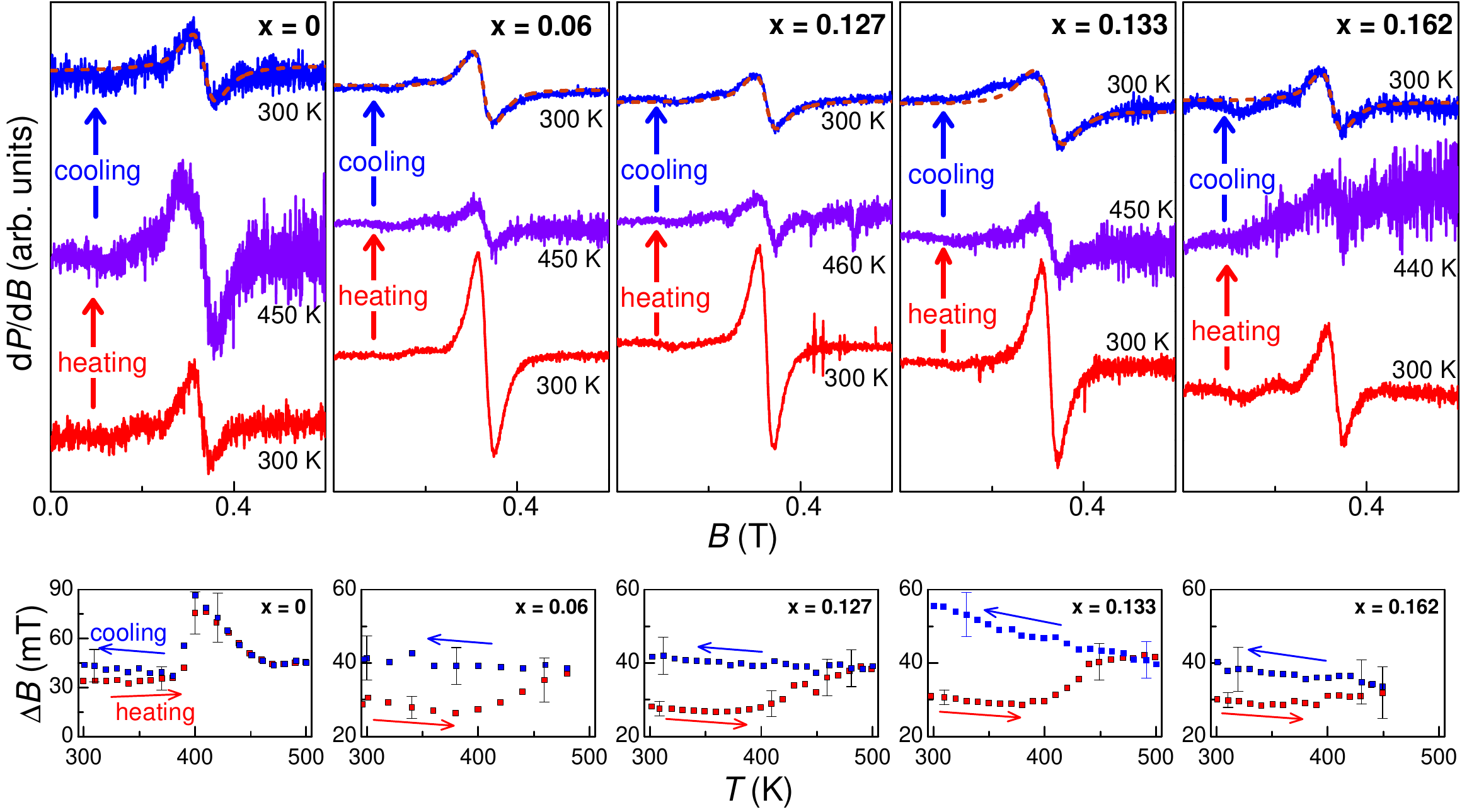}
\end{center}
\caption{
`Heat effect' on FeGa$_{3-x}$Ge$_{x}$ as seen in the temperature dependence of ESR spectra and linewidth $\Delta B(T)$. The ESR results upon ``heating'' (red arrows) and ``cooling'' (blue arrows) are reversible for undoped FeGa$_{3}$ and irreversible for the Ge-doped FeGa$_{3-x}$Ge$_{x}$. Spectra fitting by Lorentzians is shown for the 300~K spectra by red dashed lines.
}
\label{FeGa3GeHeat}
\end{figure}
Neutron scattering measurements on FeGa$_{3}$ suggested antiferromagnetic order at temperatures above 300~K \cite{gamza14a}. We therefore investigated the ESR properties of FeGa$_{3-x}$Ge$_x$ for temperatures between 300~K up to 500~K. It is also in this temperature region where upon heating the electrical resistivity starts to decrease (activated behavior with gap $\approx 0.4$~eV, Refs.\cite{wagner-reetz14c,gamza14a}) and the susceptibility shows a continuous increase \cite{gamza14a,tsujii08a}. 

By heating the FeGa$_{3-x}$Ge$_x$ samples up to 500~K and cooling them back to 300~K we observe a remarkable ``heat effect'' on the ESR spectra. As shown in figure \ref{FeGa3GeHeat} this heating-cooling cycle irreversibly changes the ESR of Ge-containing FeGa$_{3-x}$Ge$_x$ whereas the ESR properties of pure FeGa$_{3}$ can be fully recovered. The ESR properties of Ge-containing FeGa$_{3-x}$Ge$_x$ seem to change upon heating towards properties which are very similar to those of pure FeGa$_{3}$: the intensity as well as the linewidth approach values of FeGa$_{3}$.  This observation allows the conclusion that Ge doping strenghtens the high-temperature ESR signal in FeGa$_{3}$. 
However, the ESR properties of FeGa$_{3-x}$Ge$_x$ below 40~K (i.e. the CESR) are not influenced at all from such a heating-cooling cycle. This in turn means, that the Ge-ions have two possibilities to influence the ESR-probed magnetism. They contribute to a stable, heat-insensitive CESR and also to the ESR with properties as seen in the undoped FeGa$_{3}$.

\section{Discussion}

The ESR results in FeGa$_{3-x}$Ge$_x$ show a remarkable temperature behavior which strongly depends on the Ge content. The picture that evolves is characterized by a conduction spin resonance (CESR) developing at low temperatures and a local spin resonance at elevated temperatures.

Below temperatures of $\approx 40$~K and for $x>0.06$ we observed a well-defined and strong resonance in a conductive environment with typical asymmetric lineshapes. The resonant spinsystem may be interpreted as a CESR with the properties of a strongly coupled 3d -- conduction electron system. 
The spin dynamics of this coupled system can be related to the resistivity and magnetization yielding information on the nature of interactions as was done for various other itinerant 3d ferromagnetic materials \cite{shaltiel88a}.
%   
% relation to the resisitivity: no feature in \rho near 40K
%\textit{the following should be mentioned for interpreting the linewidth relation to resistivity and magnetization}
%

Two relaxation mechanisms for CESR in itinerant magnets are important namely the spin-lattice relaxation by spin-orbit coupling to transport collisions (characterized by the momentum scattering time $\tau$ which also determines the electrical resistivity) \cite{elliott54a} and a reduction of the relaxation rate by exchange enhancement (characterized by the magnetization) \cite{fulde68a}. Whether or not the temperature dependence of $\tau$ dominates the relaxation depends on how strong the exchange effect depends on temperature:
the corrections introduced from the resistivity may be small compared to line width variations due to changes in magnetization, as observed in ZrZn$_{2}$ \cite{walsh70a,forster10a} or in Sc$_{3.12}$In \cite{dunifer70a}. In TiBe$_{2}$ the linewidth broadening can be interpreted as a variation in $\tau$ because there the exchange effect is constant also at high temperatures (only d-electron pockets in the Fermi surface contribute and the interaction strength to other electrons is small).
In our case of FeGa$_{3-x}$Ge$_x$ the low-temperature relaxation is clearly related to the inverse magnetization, see figure \ref{FeGa3GedHRhoChi}. Its relation to the electrical resistivity, see inset figure \ref{FeGa3GedHRhoChi}, points to a typical CESR property although the temperature independent part of the resistivity is considerable.

%
%(Ge occupies preferentially the Ga1 site; Munevar: Ga2 is preferred)
%two sites of Fe, one is magnetic;\cite{haldolaarachchige13a}
%Ge goes to Ga, on Ga1 site;

%Fig2: background was subtracted which influences the region around 40K - better fitting results.
%argumentation for non-sequence : Ge distribution is creating at different concentration different magntic centers...  

Above temperatures of $\approx 40$~K the resonance properties point to a local spin probe although pure FeGa$_{3}$ is expected to be non-magnetic \cite{hadano09a}. Preformed magnetic moments may arise from the presence of Fe-dimers in the structure \cite{wagner-reetz14c} because their magnetic properties are very sensitive to the electronic environment \cite{haldolaarachchige13a} (being in turn affected by structural defects). Substituting some Ga by Ge stabilizes this magnetism and the high-temperature ESR spectra gain intensity without strongly changing their other parameters (see figure \ref{FeGa3GeSpecsPara}).  
Approaching the highest temperatures (500~K) the local spin resonance for FeGa$_{3-x}$Ge$_x$ with $x>0$ irreversibly changes its properties in such a way that the ESR properties of pure FeGa$_{3}$ are ``recovered''. This observation implies that heating up to 500~K changes the way Ge-doped electrons are distributed in the lattice. As there are two different Ge-substitutable Ga sites \cite{majumder16a,munevar17a} it might be possible that heat shifts the Ge to a preferred site and this site would then have less influence on the Fe-dimer magnetism. The insensitivity of the low-temperature CESR to a heating-cooling cycle may indicate that the presence of itinerant spin probes does not depend on which of the two Ga sites are substituted by Ge.

% Ge doping controls the ESR intensity also for T>40K; it seems that at low Ge concentration the signal becomes more clear - this conclusion comes also from the heating experiments. That means for the sequence of resonance fields at a given T: it depends on the Ge concentration, i.e. after heating the resonance fields at a given T should be independent on Ge-concentration 

\section{Conclusion}

The presented ESR studies confirm the rather complex microscopic formation of magnetism
%the rather complex way of magnetism formation in 
in FeGa$_{3-x}$Ge$_x$. Structural defects easily lead to residual Fe-moments which are the origin of the local-type ESR observed for temperatures above $\approx 40$~K. 
%However, there is a scenario where Fe-moments are either in an AFM ground state or in a dimerized state. Magnetic correlations are evidenced so far from NQR, uSR, magnetization, specific heat.
%magnetic susceptibility is small (0.4% Fe 0.1uB) and is strongly increasing below 40K; 
%two environments of Fe? dumbell- no dumbell?\\
%tiny amount of Fe on a site where it should not be; Ge on Fe site? (J. Grin);\\
%
The ESR properties strongly depend on the substitution of Ga sites by Ge which leads to electron doping. This supports an itinerant type of resonance below $\approx 40$~K but it also leads to a stabilization of the local Fe-dimer magnetism, increasing the intensity of the local type of resonance.

The itinerant-type resonance can be interpreted as a conduction spin resonance (CESR) of a system with a strong coupling between 3d- and conduction electrons. Strong ferromagnetic correlations allow an exchange enhancement of the spin lifetime, thus leading to a narrow and well-observable CESR. This resonance should be an appropriate tool to investigate the evolution of ferromagnetic correlations close to the suggested ferromagnetic quantum critical point \cite{majumder16a} in FeGa$_{3-x}$Ge$_x$.
 
% Recently this was similarly realized for Cr$_{2}$B where a paramagnetic to ferromagnetic transition is induced by Fe-doping \cite{arcon16a}.\\

%\emph{(the availability of high-quality single crystals is an essential precondition)}
%Two further Ge concentrations are planned to extend the ESR data of FeGa$_{3-x}$Ge$_x$: one concentration below $x=0.1$ and one above $x=0.15$. Moreover, as the properties of semiconductors strongly depend on the exact composition (which controls the presence of in-gap states or n-type or p-type doping) the investigation of further samples of pure FeGa$_{3}$ would be desirable. This is also important to confirm the very similar high-temperature ESR properties of pure and Ge doped FeGa$_{3}$. Supplementary to these planned ESR experiments the same samples should be characterized by measurements of susceptibility, resistivity, and specific heat.\\
\section*{Acknowledgements} We acknowledge valuable discussions with Raul Cardoso-Gil, Dmitry Sokolov and Hiroshi Yasuoka. 
%B.K. acknowledges financial support from the Max Planck POSTECH/Korea Research Initiative. 
B.K. was supported by the National Research Foundation of Korea funded by the Korean government (MSIT) (No.2016K1A4A4A01922028).

\section*{References}
\bibliography{JoergBib}

\providecommand{\newblock}{}
\begin{thebibliography}{10}
\expandafter\ifx\csname url\endcsname\relax
  \def\url#1{{\tt #1}}\fi
\expandafter\ifx\csname urlprefix\endcsname\relax\def\urlprefix{URL }\fi
\providecommand{\eprint}[2][]{\url{#2}}

\bibitem{yin10a}
Yin Z~P and Pickett W~E 2010 {\em Phys. Rev. B\/} {\bf 82} 155202

\bibitem{arita11a}
Arita M, Shimada K, Utsumi Y, Morimoto O, Sato H, Namatame H, Taniguchi M,
  Hadano Y and Takabatake T 2011 {\em Phys. Rev. B\/} {\bf 83} 245116

\bibitem{majumder16a}
Majumder M, Wagner-Reetz M, Cardoso-Gil R, Gille P, Steglich F, Grin Y and
  Baenitz M 2016 {\em Phys. Rev. B\/} {\bf 93} 064410

\bibitem{singh13b}
Singh D~J 2013 {\em Phys. Rev. B\/} {\bf 88} 064422

\bibitem{alvarez-quiceno16a}
Alvarez-Quiceno J~C, Cabrera-Baez M, Ribeiro R~A, Avila M~A, Dalpian G~M and
  Osorio-Guill\'en J~M 2016 {\em Phys. Rev. B\/} {\bf 94} 014432

\bibitem{munevar17a}
Munevar J, Cabrera-Baez M, Alzamora M, Larrea J, Bittar E~M, Baggio-Saitovitch
  E, Litterst F~J, Ribeiro R~A, Avila M~A and Morenzoni E 2017 {\em Phys. Rev.
  B\/} {\bf 95}(12) 125138

\bibitem{krellner08a}
Krellner C, F\"orster T, Jeevan H, Geibel C and Sichelschmidt J 2008 {\em Phys.
  Rev. Lett.\/} {\bf 100} 066401

\bibitem{walsh70a}
Walsh W~M, Knapp G~S, L~W~Rupp J and Schmidt P~H 1970 {\em J. Appl. Phys.\/}
  {\bf 41} 1081

\bibitem{forster10a}
F\"orster T, Sichelschmidt J, Gr\"uner D, Brando M, Kimura N and Steglich F
  2010 {\em J. Phys.: Conf. Ser.\/} {\bf 200} 012035

\bibitem{shaltiel88a}
Shaltiel D 1988 {\em Helv. Phys. Acta\/} {\bf 61} 505

\bibitem{shaltiel80a}
Shaltiel D, Monod P and Felner I 1980 {\em J. Physique Lett.\/} {\bf 41} 567

\bibitem{rauch15a}
Rauch D, Kraken M, Litterst F~J, S\"ullow S, Luetkens H, Brando M, F\"orster T,
  Sichelschmidt J, Neubauer A, Pfleiderer C, Duncan W~J and Grosche F~M 2015
  {\em Phys. Rev. B\/} {\bf 91} 174404

\bibitem{arcon16a}
Arcon D, Schoop L~M, Cava R~J and Felser C 2016 {\em Phys. Rev. B\/} {\bf 93}
  104413

\bibitem{wykhoff07b}
Wykhoff J, Sichelschmidt J, Lapertot G, Knebel G, Flouquet J, Fazlishanov I~I,
  Krug~von Nidda H~A, Krellner C, Geibel C and Steglich F 2007 {\em Science
  Techn. Adv. Mat.\/} {\bf 8} 389

\bibitem{wagner-reetz14c}
Wagner-Reetz M, Kasinathan D, Schnelle W, Cardoso-Gil R, Rosner H, Grin Y and
  Gille P 2014 {\em Phys. Rev. B\/} {\bf 90}(19) 195206

\bibitem{abragam70a}
Abragam A and Bleaney B 1970 {\em Electron Paramagnetic Resonance of Transition
  Ions\/} (Oxford: Clarendon Press)

\bibitem{gamza14a}
Gam{\.z}a M~B, Tomczak J~M, Brown C, Puri A, Kotliar G and Aronson M~C 2014
  {\em Physical Review B\/} {\bf 89} 195102

\bibitem{mendez15a}
Mendez J~H, Ekuma C~E, Wu Y, Fulfer B~W, Prestigiacomo J~C, Shelton W~A,
  Jarrell M, Moreno J, Young D~P, Adams P~W, Karki A, Jin R, Chan J~Y and
  DiTusa J~F 2015 {\em Phys. Rev. B\/} {\bf 91} 144409

\bibitem{cox87a}
Cox D~L 1987 {\em Phys. Rev. Lett.\/} {\bf 58} 2730

\bibitem{barnes81a}
Barnes S~E 1981 {\em Adv. Phys.\/} {\bf 30} 801

\bibitem{mishra98a}
Mishra S~G and Sreeram P~A 1998 {\em Phys. Rev. B\/} {\bf 57} 2188

\bibitem{tsujii08a}
Tsujii N, Yamaoka H, Matsunami M, Eguchi R, Ishida Y, Senba Y, Ohashi H, Shin
  S, Furubayashi T, Abe H and Kitazawa H 2008 {\em J. Phys. Soc. Jap.\/} {\bf
  77} 024705

\bibitem{elliott54a}
Elliott R~J 1954 {\em Phys. Rev.\/} {\bf 96} 266

\bibitem{fulde68a}
Fulde P and Luther A 1968 {\em Phys. Rev.\/} {\bf 175} 337

\bibitem{dunifer70a}
Dunifer G~L, Knapp G~S and Corenzwit E 1970 {\em J. Appl. Phys.\/} {\bf 41}
  1075

\bibitem{hadano09a}
Hadano Y, Narazu S, Avila M~A, Onimaru T and Takabatake T 2009 {\em J. Phys.
  Soc. Jap.\/} {\bf 78} 013702

\bibitem{haldolaarachchige13a}
Haldolaarachchige N, Prestigiacomo J, Phelan W~A, Xiong Y~M, McCandless G, Chan
  J~Y, DiTusa J~F, Vekhter I, Stadler S, Sheehy D, Adams P and Young D 2013
  {\em arXiv\/} {\bf 1304} 1897

\end{thebibliography}

\end{document}